\documentstyle[12pt,epsf]{article}

\makeatletter
\@addtoreset{equation}{section}
\@addtoreset{footnote}{page}
\makeatother

\newcommand{\bra}[1]{\left\langle\, #1\,\right|}
\newcommand{\ket}[1]{\left|\, #1\,\right\rangle}
\newcommand{\sbra}[1]{\left\langle #1\right|}
\newcommand{\sket}[1]{\left| #1\right\rangle}
\newcommand{\VEV}[1]{\left\langle #1\right\rangle}

\newcommand{\Bracket}[2]{\VEV{#1\,\Big|\,#2}}
\newcommand{\wt}{\widetilde}
\newcommand{\wh}{\widehat}
\newcommand{\ol}{\overline}
\newcommand{\del}{\partial}
\newcommand{\ra}{\rightarrow}

\newcommand{\nn}{\nonumber}
\newcommand{\half}{\frac{1}{2}}

\def\tr{\mathop{\rm tr}\nolimits}
\def\Tr{\mathop{\rm Tr}\nolimits}

\newcommand{\gym}{g_{Y\!M}}
\newcommand{\cA}{{\cal A}}

\newcommand{\R}{{\bf R}}
\newcommand{\C}{{\bf C}}
\newcommand{\Z}{{\bf Z}}

\def\matt[#1,#2,#3,#4]{\left(%
\begin{array}{cc} #1 & #2 \\ #3 & #4 \end{array} \right)}

\def\v2#1{\vv2[#1]}
\def\vv2[#1,#2]{\left(%
{#1 \atop #2}\right)}

\begin{document}

\begin{titlepage}
\vspace*{-2.0cm}
\null
\begin{flushright}
hep-th/0403247 \\
YITP-04-13\\
March, 2004
\end{flushright}
\vspace{0.5cm}
\begin{center}
{\Large \bf
QED and String Theory\\
}
\lineskip .75em
\vskip1.5cm
\normalsize

{\large
Shigeki Sugimoto\footnote{
E-mail:\ \ {\tt sugimoto@yukawa.kyoto-u.ac.jp} } and
Kazuyoshi Takahashi \footnote{
E-mail:\ \ {\tt kazuyosi@yukawa.kyoto-u.ac.jp} }
}
\vskip 2.5em

{
\it
Yukawa Institute for Theoretical Physics,
Kyoto University,\\
Kyoto 606-8502, Japan
}
\vskip 1em
\vskip 2em

{\bf Abstract}

\end{center}

We analyze the $D9$-$\ol{D9}$ system in type IIB string theory 
using $Dp$-brane probes.
It is shown that the world-volume theory of the probe $Dp$-brane
contains two-dimensional and four-dimensional QED
in the cases with $p=1$ and $p=3$, respectively,
and some applications of the realization of these well-studied
quantum field theories are discussed.
In particular, the two-dimensional QED (the Schwinger model)
is known to be a solvable theory
and we can apply the powerful field theoretical techniques,
such as bosonization, to study the D-brane dynamics. 
The tachyon field created by the $D9$-$\ol{D9}$ strings 
appears as the fermion mass term in the Schwinger model and
the tachyon condensation is analyzed by using the bosonized description.
In the T-dualized picture, we obtain the potential between
a $D0$-brane and a $D8$-$\ol{D8}$ pair using the Schwinger model
and we observe that it consists of the energy carried by fundamental
strings created by the Hanany-Witten effect and the vacuum energy
due to a cylinder diagram. The $D0$-brane is treated
quantum mechanically as a particle trapped in the potential,
which turns out to be a system of a harmonic oscillator.

As another application, we obtain a matrix theory description of QED
using Taylor's T-duality prescription,
which is actually applicable to a wide variety of
field theories including the realistic QCD.
We show that the lattice gauge theory is naturally obtained
by regularizing the matrix size to be finite.

\end{titlepage}

\baselineskip=0.7cm


\section{Introduction}

The interplay between string theory and quantum field theory
has been one of the most successful subject during the
second revolution of string theory.
There are many things that we can learn from it.
For example, quantum field theory often provides useful
tools to study non-perturbative aspects of string theory.
Even though the non-perturbative formulation of string theory
is not available yet, we can analyze non-perturbative effects
using the techniques developed in the quantum field theory
once we know the realization of the quantum field theory
in string theory. On the other hand,
we can apply various string duality (such as S-duality, T-duality,
M-theory, open/closed duality, etc.) to quantum field theory.
If we are lucky enough, we would be able to obtain
a new description of the quantum field theory.
However, most of the works along this line is done in
supersymmetric situations. Since our goal is to understand
the real world, it would be quite important to investigate
non-supersymmetric models.

One of the purpose of this paper is to analyze 
unstable D-brane systems\footnote{
See \cite{Se} for a review.} by using probe D-branes.
As a typical example, we consider
the $D9$-$\ol{D9}$ system in type IIB string theory 
and take a $Dp$-brane ($p=1,3$) as a probe.\cite{Ho} \
As we will explain in section \ref{twoqed},
the world-volume theory on the $Dp$-brane contains
$(p+1)$ dimensional QED.

The realization of the four dimensional QED in string theory
could be interesting since it is a realistic system.
It would be interesting if we could say something realistic
using string theory,
though we will not consider much about it in this paper.

In the $p=1$ case, we obtain the two dimensional QED, which is often
called as the Schwinger model. The Schwinger model is known to be
one of the exactly solvable interacting
quantum field theories.\cite{Schwinger} \
Actually,
it has been shown that the system is equivalent to a free massive scalar
field theory by using bosonization techniques.\cite{Low} \
We can thus analyze the D-brane dynamics using the field theoretical
results in the Schwinger model.
Being two-dimensional, it is not a realistic model, however
there are many features common to the four dimensional QCD
(such as confinement, chiral symmetry breaking, axial anomaly,
instantons, $\theta$-vacuum etc.)
which make this theory even more interesting.

The $D9$-$\ol{D9}$ system is an unstable D-brane system and it is known
that there is a tachyon field created by the open string stretched
between the $D9$-brane and the $\ol{D9}$-brane. When the tachyon
field homogeneously condenses, the $D9$-$\ol{D9}$ pairs
are believed to be annihilated.\cite{Sen1,Sen2,Se} \
 More interestingly,
when the tachyon field takes a vortex configuration,
it represents a $D7$-brane.\cite{Senv,Se} \
The lower dimensional
D-branes can be similarly constructed by non-trivial
tachyon configurations in the $D9$-$\ol{D9}$ system.\cite{Wittenv}
When we put the D-brane probe in this system, the tachyon field
is interpreted as the fermion mass parameter in the world-volume
theory. We will observe these phenomena in terms of the
bosonized description of the Schwinger model.

When we compactify the direction parallel to the D-brane probe
to a torus, we can T-dualize the system to obtain a lower
dimensional description of QED. The T-duality prescription
given in \cite{Taylor1,Banks,Ganor,Taylor2}
can also be applied to the world-volume
theory on the D-brane in the presence of the $D9$-$\ol{D9}$ pairs.
Actually, since the essential step in the prescription is just
the Fourier transformation, it is applicable to any field theory
compactified on a torus. By T-dualizing all the space-time
directions, we obtain a matrix theory description.
The size of the matrix variables here is infinite, since
there are infinitely many copies of the D-branes in the covering space.
We shall show that when we regularize the matrix size to be finite
in this matrix theory description, we naturally obtain usual
lattice gauge theory.

The paper is organized as follows.
First we summarize the world-volume theory of
the $Dp$-brane probe in the $D9$-$\ol{D9}$ system
and see how QED is realized on it
in section \ref{twoqed}.
In section \ref{qedbrane}, we
study D-brane dynamics using the Schwinger model.
After a brief review of the bosonization,
we consider tachyon condensation in terms of the
bosonized description of the Schwinger model.
In section \ref{D0}, we T-dualize the spatial direction
of the $D1$-brane probe and consider a $D0$-brane in the presence of
a $D8$-$\ol{D8}$ pair.
T-duality is more systematically considered in section \ref{Matrix},
in which the relation between the matrix regularization
and the lattice theory is discussed.
Section \ref{discuss} is devoted to discussion
and we make some speculation on the S-duality
of the $D9$-$\ol{D9}$ system, which was actually our
first motivation for the present work.

\section{QED in String Theory}
\label{twoqed}

In this section, we consider
the $D9$-$\ol{D9}$ system in type IIB string theory 
and take a $Dp$-brane ($p$: odd) as a probe.\footnote{This
configuration
is first analyzed by K. Hori in \cite{Ho}.}
We summarize the world-volume theory of the D$p$-brane probe
fixing our notation.
We will soon show that the world-volume theory of the probe D$p$-brane
contains two-dimensional and four-dimensional QED
in the cases with $p=1$ and $p=3$, respectively.

\subsection{$Dp$-branes in the Type IIB $D9$-$\ol{D9}$ System}
Let us consider $n$ $Dp$-branes extended along $x^0,\dots,x^p$
directions in the presence of background $N$
$D9$-brane - $\ol{D9}$-brane pairs.
The world-volume fields on the $Dp$-branes consist of those
created by the $p$-$p$ strings, $p$-$9$ strings and
$p$-$\ol 9$ strings, which are the open strings
with ends on the respective D-branes.

The massless fields generated by the $p$-$p$ strings are
the same as those obtained by the dimensional reduction
of $10$ dimensional $U(n)$ super Yang-Mills theory, namely,
a gauge field $A_{\mu}$ ($\mu = 0, \dots ,p$),
scalar fields $\Phi^{i}$ ($i=p+1, \cdots , 9$)
and fermions $S$. These fields transform as the adjoint representation
of the gauge group $U(n)$.

As for the $p$-$9$ strings and $p$-$\ol 9$ strings,
the mass of the lowest modes depends on $p$.
In fact, in the mass shell condition
$L_0\ket{\mbox{phys}}=a\ket{\mbox{phys}}$,
the zero-point energy $a$ is given by
\begin{eqnarray}
a^{\rm R}=0,~~~~~
a^{\rm NS}=(p-1)\left(-\frac{1}{24}-\frac{1}{48}
 \right)+(9-p)\left(\frac{1}{48}+\frac{1}{24}
 \right)=\frac{5-p}{8}
 \label{aNS}
\end{eqnarray}
for R-sector and NS-sector, respectively.
\cite{PoChJo} \
As we can see from (\ref{aNS}),
the lowest mass states in the NS-sector are massive for $p=1,3$, massless
for $p=5$ and tachyonic for $p=7$. From now, we shall concentrate on
the $p=1,3$ cases, in which we can simply forget about the 
extra massless or tachyonic bosons
coming from the $p$-$9$ and $p$-$\ol 9$ strings
in the low energy physics.

On the other hand,
since the normal ordering constant for the R-sector $a^R$ is
always zero, we have massless fermions.
Note that the world-sheet fermions $\psi^{i}$ ($i=p+1,\dots,9$)
which correspond to the directions transverse to the $Dp$-brane
do not have zero modes,
and hence the massless states are generated by the 
zero modes of $\psi^{\mu}$ ($\mu=0,\dots,p$).
Then, the ground states in the R-sector consist of
positive and negative chirality spinor representations of
the Lorentz group $SO(1,p)$.
One of these two spinors is removed by GSO projection.
We choose the positive chirality spinor as physical states
created by the $p$-$9$ strings
and the corresponding fields are denoted as $\lambda_+^I$,
where $I=1,\dots,N$ labels the Chan-Paton indices of the $D9$-branes.
These fermion fields are considered as complex fermions, since
the $p$-$9$ string have two orientations.
The massless fermion fields generated by the $p$-$\ol 9$ strings
can be obtained in a similar way. The only point we should
notice is that the GSO projection for the $p$-$\ol 9$
strings is opposite to the one we chose for
the $p$-$9$ strings.
Therefore, the massless fermion fields
 coming from the
$p$-$\ol 9$ strings belong to the negative chirality spinor
representation of the Lorentz group $SO(1,p)$
and these fields are denoted by $\lambda_-^{\bar I}$.
Here $\bar{I}=1,\dots,N$ labels the Chan-Paton indices of the
$\ol{D9}$-branes.

In summary,
the massless fields on the $Dp$-branes for $p=1,3$
are as listed in Table \ref{p=1} and Table \ref{p=3}, respectively.
\begin{table}[htb]
\begin{center}
$$
\begin{array}{c|cccc}
\hline\hline
\mbox{field} & U(n) & SO(1,1) & SO(8) & U(N)\times U(N) \\
\hline
A_{\mu} & \textbf{adj.} & 2 & 1 & (1,1)\\
\Phi^{i}  & \textbf{adj.} & 1 & 8 & (1,1)\\
S_+ & \textbf{adj.} & 1_+ & 8_+ & (1,1)\\
S_- & \textbf{adj.} & 1_- & 8_- & (1,1)\\
\hline
\lambda_{+}^{I}  & \textbf{fund.} &  1_+
 &1& (\textbf{fund.},1) \\
\lambda_{-}^{\ol{I}}  & \textbf{fund.}  & 1_-
&1& (1,\textbf{fund.})\\
\hline
\end{array}
$$
\parbox{75ex}{
\caption{
\small
The massless fields on the $D1$-brane in the presence
of $D9$-$\ol{D9}$ pairs. Here
 $1_+$ and $1_-$ denote the
positive and negative chirality Majorana-Weyl spinor
representation
of the Lorentz group $SO(1,1)$, respectively. Similarly,
 $8_+$ and $8_-$ denote the
positive and negative chirality Majorana-Weyl spinor
representation of the $SO(8)$, respectively.
\textbf{adj.} and \textbf{fund.} represent the adjoint and
fundamental representations, respectively.
}
\label{p=1}
}
\end{center}
\end{table}
\begin{table}[htb]
\begin{center}
$$
\begin{array}{c|cccc}
\hline\hline
\mbox{field} & U(n) & SO(1,3) & SO(6) & U(N)\times U(N) \\
\hline
A_{\mu} & \textbf{adj.} & 4 & 1 & (1,1)\\
\Phi^{i}  & \textbf{adj.} & 1 & 6 & (1,1)\\
S & \textbf{adj.} & 2_+ & 4_+ & (1,1)\\
\hline
\lambda_{+}^{I}  & \textbf{fund.} & 2_+
 &1& (\textbf{fund.},1) \\
\lambda_{-}^{\ol{I}}  & \textbf{fund.}  & 2_-
&1& (1,\textbf{fund.})\\
\hline
\end{array}
$$
\parbox{75ex}{
\caption{
\small
The massless fields on the $D3$-brane in the presence
of $D9$-$\ol{D9}$ pairs. Here
 $2_+$ and $2_-$ denote the
positive and negative chirality spinor representation
of the Lorentz group $SO(1,3)$. Similarly,
 $4_+$ denotes the
positive chirality spinor representation of the $SO(6)$.
}
\label{p=3}
}
\end{center}
\end{table}
The $U(N)\times U(N)$ symmetry on the Tables \ref{p=1} and \ref{p=3}
is the gauge symmetry of the $N$ $D9$-$\ol {D9}$ pairs
which can be seen as the global symmetry on the $Dp$-brane
world-volume. Note that the anomaly with respect to
the gauge symmetry $U(n)$ on the world-volume precisely
cancels if and only if the number of the $D9$-branes and 
the $\ol{D9}$-branes are the same. This condition is also
consistently required from the cancellation of the RR-tadpole
or the 10 dimensional gauge anomaly in the $D9$-$\ol {D9}$
system.\cite{Sr,ScWi} \
It is sometimes useful to combine $\lambda_+$ and $\lambda_-$
to make Dirac fermions
\begin{eqnarray}
\lambda^{I}=\v2{\lambda_+^{I},\lambda_-^{I}},~~~~(I=1,\dots,N),
\label{Dirac}
\end{eqnarray}
though only the diagonal $U(N)$ component of the $U(N)\times U(N)$
symmetry becomes manifest in this notation.

When $n=1$, the world-volume
theory of the $Dp$-brane becomes very simple,
since the gauge group $U(n)$ is Abelian.
Then, the low energy world-volume action is
\begin{eqnarray}
S_{Dp}
=\int\! d^{p+1}x\left\{
-\frac{1}{4 \gym^2} F_{\mu \nu }F^{\mu \nu}
+i\ol{\lambda}_I\gamma^{\mu}D_{\mu}\lambda^I
+\frac{1}{2}\partial_{\mu}\Phi^{i}\partial^{\mu}\Phi^{i}
+i\ol{S}\gamma^{\mu}\partial_{\mu}S
\,\right\},
\label{Dpaction}
\end{eqnarray}
where $D_\mu=\del_\mu+iA_\mu$ is the covariant derivative.
The gauge coupling $\gym$ is related to the string coupling $g_s$ and
the string length $l_s$ by
\begin{equation}
\gym^2=\frac{g_s}{l_{s}}(2 \pi l_{s})^{p-2}.
\end{equation}
We take the limit $l_s\ra 0$ keeping $\gym$ fixed, in which
higher order terms in the action (\ref{Dpaction}) as well as
the coupling to the bulk fields are suppressed.
In the action (\ref{Dpaction}), the $U(1)$ gauge field
and the Dirac fermion $\lambda^I$ make QED with $N$ flavors,
while the fields $\Phi^i$ and
$S$ are decoupled from this sector.

\subsection{Turning on Bulk Fields}
\label{Bulk}

It is also interesting to see what happens if we turn on the
bulk fields as a background. 
When we turn on the R-R fields, we should take into account the
Chern-Simons term
\begin{eqnarray}
S_{CS}=\mu_p \int_{p+1} C\wedge \tr\left(e^{2\pi\alpha' F}\right),
\end{eqnarray}
where $C=C_0+C_2+\dots$ is a formal sum of the R-R $k$-form fields
$C_k$.\cite{Li,Douglas,GrHaMo}
In particular, the R-R $0$-form field $C_0$ acts as a theta
parameter in the $Dp$-brane world-volume theory.
For example, in the case with $p=1$, we have
\begin{eqnarray}
S_{C_0}=\int C_0\, F,
\label{C0CS}
\end{eqnarray}
which will play an important role in the following sections.

The tachyon field $T$ created by $9$-$\ol{9}$ strings
will couple to the fermions $\lambda^I$ as
\begin{eqnarray}
S_T=\int\! d^{p+1}x\,\,
T^I_{~\bar I}\,\lambda^*_{+I} \lambda_-^{\bar I}+\mbox{h.c.},
\label{ST}
\end{eqnarray}
which behaves as a mass term for the fermions
when the tachyon field is non-zero.\footnote{
Here we assume that the possible $T$
dependence in the kinetic term of the fermion $\lambda^I$
can be absorbed by the redefinition of the fermion fields.
More general $T$ dependence in (\ref{ST}) such as
$S_T=\int\! d^{p+1}x\,\,
f(T)^I_{~\bar I}\,\lambda^*_{+I} \lambda_-^{\bar I}+\mbox{h.c.}$
is also assumed to be absorbed by the redefinition
of the tachyon field
$f(T)\ra T$. The $T$ dependence of the kinetic term of the
gauge field can appear as the higher loop corrections
which will vanish in the decoupling limit for $p=1$.
}
As observed in \cite{Ho}, when the tachyon get large vev,
the fermions $\lambda^I$ will become massive
and decouple from the low energy degrees of freedom,
which is consistent with the annihilation of the $D9$-$\ol{D9}$
pairs via the tachyon condensation.\cite{Sen1,Sen2,Se} \

When we turn on the bulk gauge fields 
$\cA_M$ and $\ol \cA_M$ which correspond to the $U(N)\times U(N)$
gauge symmetry of the $D9$-$\ol{D9}$ system,
we should add
\begin{eqnarray}
S_\cA=-\int\! d^{p+1}x\,\,
\ol\lambda_{I}\gamma^\mu
(\cA^+_\mu+\gamma_5\cA^-_\mu)^I_{~J}\lambda^J,
\label{D9gauge}
\end{eqnarray}
where $\cA^\pm_\mu=\half(\cA_\mu+\cA_i\del_\mu X^i
\pm (\ol\cA_\mu+\ol\cA_i\del_\mu X^i))$,
($X^i\equiv T_p^{-1/2}\Phi^i$ where $T_p$ is the $Dp$-brane
tension).
Therefore $\cA_\mu^+$ and  $\cA_\mu^-$ can be regarded as
sources which couple to the vector and axial currents
$j^\mu=\ol\lambda\gamma^\mu\lambda$ and
$j_5^\mu=\ol\lambda\gamma^\mu\gamma_5\lambda$,
respectively.

\section{QED${}_{2}$ and D-Brane Dynamics \label{qedbrane}}

In this section, we consider the $p=1$ case in which
we can realize the two-dimensional QED as we have seen in the
last section.
The two-dimensional massless QED (the massless Schwinger model)
is one of the exactly solvable
interacting quantum field theories.\cite{Schwinger} \
It has been shown that the system is equivalent to a free massive scalar
field theory by using bosonization
techniques.\cite{Low,Abdalla,ZinnJusting} \
Here we will discuss various field theoretical results in
the two-dimensional QED in terms of string theory
and consider some applications to the D-brane dynamics.

\subsection{The Schwinger Model and Bosonization}
\label{schwinger}

Here we briefly review some of the results
in the Schwinger model.\footnote{See  for example
\cite{Abdalla,ZinnJusting} for
a review of the Schwinger model and its bosonization.} \
We restrict our discussion to the one flavor case
in this subsection for simplicity.
The multi-flavor case will be discussed in section \ref{multi}.

The action of the (massless) Schwinger model is
\begin{equation}
S=\int\! d^2x\left\{\,
-\frac{1}{4\gym^{2}}F^{\mu \nu}F_{\mu \nu}
+i\ol{\lambda}\gamma^{\mu}(\partial_{\mu}+iA_{\mu})\lambda
\,\right\}
\label{Saction}
\end{equation}
where $\lambda=(\lambda_+,\lambda_-)^T$ is a complex Dirac fermion.
We take the representation of the gamma matrices as
$\gamma^{0}=\sigma^{1}$, $\gamma^{1}=-i\sigma^{2}$
and $\gamma^{5}=\gamma^{0}\gamma^{1}=\sigma^{3}$,
where $\sigma^{i}$ are the Pauli matrices.

The action (\ref{Saction}) is invariant under
the vector-like $U(1)_V$ and axial $U(1)_A$ transformations
\begin{eqnarray}
U(1)_V&:&\lambda\ra e^{i\alpha}\lambda,\\
U(1)_A&:&\lambda\ra e^{i\alpha\gamma_5}\lambda,
\label{U1axial}
\end{eqnarray}
though the later is anomalous.
Correspondingly, we have
the vector current $j^{\mu}=\ol{\lambda}\gamma^{\mu}\lambda$
and the axial current $j_{5}^{\mu}
=\ol{\lambda}\gamma^{\mu}\gamma^{5}\lambda$
which satisfy
\begin{eqnarray}
\del_\mu j^\mu=0,~~~~~\del_\mu j_5^\mu=\frac{1}{\pi}F_{01}.
\label{currentcons}
\end{eqnarray}

Remarkably, using the bosonization techniques,
it has been shown that the Schwinger model is equivalent to
a two-dimensional system of a free massive scalar field with a
standard Hamiltonian
\begin{eqnarray}
H=\half\int\! dx^1\left\{\,
\pi_\varphi^2+(\del_1\varphi)^2+ m^2\varphi^2
\,\right\}
\label{phiHam}
\end{eqnarray}
where $\pi_\varphi$ is the momentum conjugate to the scalar field
$\varphi$ and the mass is related to the gauge coupling by
$m^2=\gym^2/\pi$.
The scalar field  $\varphi$ and its conjugate momentum $\pi_\varphi$
is related to the field strength and axial charge density, respectively,
as
\begin{eqnarray}
\varphi=-\frac{\sqrt{\pi}}{\gym^2}F_{01},~~~
\pi_\varphi=\sqrt{\pi}\,\lambda^\dag\sigma^3\lambda.
\label{phipi}
\end{eqnarray}
Using the equation of motion
\begin{equation}
\del_\mu F^{\mu\nu}=\gym^{2}j^\nu
\end{equation}
these relations can be written in a covariant
form as
\begin{eqnarray}
j_5^\mu=\epsilon^{\mu\nu}j_\nu=\frac{1}{\sqrt{\pi}}\del^\mu\varphi,
\label{currentphi}
\end{eqnarray}
where $\epsilon^{10}=-\epsilon^{01}=+1$.
It is easy to see that the equation (\ref{currentcons}) is
consistent with the equation of motion for the free scalar field $\varphi$
under this correspondence (\ref{currentphi}).

\subsection{Axial Anomaly and the Green-Schwarz Mechanism}
\label{GS}

As we mentioned in section \ref{schwinger},
the axial $U(1)_A$ symmetry is anomalous in the Schwinger model.
This fact might sound puzzling since this axial $U(1)_A$ symmetry
is a part of the $U(1)\times U(1)$ gauge symmetry of the $D9$-$\ol{D9}$
system which has to be anomaly free as a consistent theory.
The resolution of this puzzle is obtained by a standard
anomaly inflow argument as given in \cite{GrHaMo}
which we shortly explain in the following.

The anomaly cancellation in the type IIB $D9$-$\ol{D9}$ system
was first discussed in \cite{Sr} and further developed
in \cite{ScWi}. It was shown that the anomaly is canceled by
the Green-Schwarz mechanism \cite{Green-Schwarz} which requires
the non-trivial gauge transformation rules for the R-R fields. 
In particular, the R-R $0$-form field $C_0$ transforms as
\begin{eqnarray}
C_0\ra C_0+\frac{\alpha}{\pi}
\label{shift}
\end{eqnarray}
under the axial $U(1)_A$ transformation (\ref{U1axial}).
Accordingly, the Chern-Simons term (\ref{C0CS}) will be shifted
by $\frac{\alpha}{\pi}\int F$ which precisely cancels the anomalous
transformation of the path integral measure of the fermion $\lambda$;
\begin{eqnarray}
{\cal D}\lambda\ra{\cal D}\lambda\,\exp
\left(-i\,\frac{\alpha}{\pi}\int F\right).
\end{eqnarray}
Therefore there is actually no anomaly for the axial $U(1)_A$
transformation
when we combine (\ref{U1axial}) with the shift of the R-R $0$-form
field (\ref{shift}).

\subsection{Massive Schwinger Model and Tachyon Condensation}
\label{tachyon}

When the tachyon field becomes non-zero, the mass term for the
fermion $\lambda$ is induced as we saw in section \ref{Bulk}.
The two dimensional QED with a fermion mass term
(the massive Schwinger model)
is no longer a solvable theory, but we can still use the bosonization
techniques.
Actually, using the bosonization rule
\begin{eqnarray}
\lambda_+^*\lambda_-\propto e^{i(2\sqrt{\pi}\varphi-\theta)},
\end{eqnarray}
the bosonized action
for the massive Schwinger model is obtained
as \cite{CJS,Coleman,Abdalla,ZinnJusting}
\begin{eqnarray}
S=\int\! d^2 x \left\{\,
\half\del_\mu\varphi\del^\mu\varphi-\frac{m^2}{2}\varphi^2
+c\,T \left(\cos\left(2\sqrt{\pi}\,\varphi-\theta\right)-1\right)
\right\},
\label{MSM}
\end{eqnarray}
where $m^2=\gym^2/\pi=g_s/(2\pi^2\alpha')$,
$\theta= 2\pi C_0$ is the theta parameter, $c$ is a numerical constant
and $T$ is the tachyon field appeared in (\ref{ST}),
which is set to be real and non-negative
by the axial $U(1)_A$ transformation.
Here we have added a constant term to the potential so that
there is a finite energy stable vacuum in the $T\ra\infty$ limit,
since the $D1$-brane remains stable
when the $D9$-$\ol{D9}$ pair is annihilated
via the tachyon condensation.

Let us now consider the tachyon condensation using this bosonized
action (\ref{MSM}). When the tachyon approach the minimum of the
potential, we expect that $T$ becomes very large.
As we take the limit $T\ra\infty$, the finite energy configurations
are those with
\begin{eqnarray}
\varphi=\sqrt{\pi}\left(n+\frac{\theta}{2\pi}\right),~~~(n\in\Z),
\label{varphi}
\end{eqnarray}
which means that the scalar field $\varphi$ can no longer
fluctuate.

The physical interpretation of the discrete values in (\ref{varphi})
is clear from the relation (\ref{phipi}). Since the scalar field $\varphi$
is proportional to the electric flux $F_{01}$ which induces
the fundamental string charge, it represents that the $D1$-brane
makes a bound state with $n$ fundamental strings, which is called
the $(n,1)$ string.

Then, the energy density ${\cal E}$
carried by the configuration (\ref{varphi}) is
\begin{eqnarray}
{\cal E}=
\frac{m^2}{2}\pi \left(n+\frac{\theta}{2\pi}\right)^2
=\frac{g_s}{4\pi\alpha'}(n+C_0)^2.
\label{ene}
\end{eqnarray}
This result can be
reproduced by using the tension formula for the $(n,1)$ string.
In fact, the tension of the $(n,1)$ string
is given as \cite{Schwarz,Wittenb}
\begin{eqnarray}
T_{(n,1)}=\frac{1}{2\pi\alpha'}\sqrt{(n+ C_0)^2+\frac{1}{g_s^2}},
\end{eqnarray}
and hence the excess energy density of the system is 
\begin{eqnarray}
T_{(n,1)}-T_{D1}=\frac{1}{2\pi\alpha' g_s}\left(
\sqrt{1+g_s^2(n+C_0)^2}-1
\right)\ra \frac{g_s}{4\pi\alpha'}(n+C_0)^2,
\end{eqnarray}
which agrees with (\ref{ene})
in the limit $g_s\ra 0$ with $\gym^2=g_s/(2\pi\alpha')$ fixed.

It is interesting to note that we have obtained the
correct quantization condition for the electric flux
using our bosonized description of the $D1$-brane.
 The quantization of the electric
flux is rather non-trivial if we use the DBI action to describe
the system. (See \cite{Wittenb,Taylor2})
The action (\ref{MSM}) also tells us that
when $T$ is finite, which corresponds to the situation that
the tachyon field is not at the minimum of the potential,
the scalar field $\varphi$ can fluctuate and hence the electric flux
is not fixed to the discrete values. Namely, the $n$-units
of the fundamental string in the $(n,1)$ string is dissolved
in the bulk. This is possible because the fundamental strings
are unstable in the background of $D9$-$\ol{D9}$ pairs,
since they can decay to short pieces of open strings ending
at the $D9$ or $\ol{D9}$-branes.

\subsection{Kinks and Vortices}

Since the fermion mass is given by the tachyon field in our set up,
it can vary along the $D1$-brane world-sheet.
As an example,
let us consider a kink-like tachyon configuration such as
\begin{eqnarray}
T\sim x^1,
\label{kink}
\end{eqnarray}
where $x^1$ is the spatial direction along the $D1$-brane.
In this case, since the tachyon field becomes large at $x^1\ra\pm\infty$,
the scalar field $\varphi$ takes discrete values as (\ref{varphi})
at the spatial infinity. An interesting point here is that
we can consider a kink configuration for the scalar field $\varphi$.
Namely, the integer $n$ in (\ref{varphi}) chosen for $x^1\ra -\infty$
can be different from that for $x^1\ra +\infty$.
This means that the fundamental string ingredients
can escape from the $D1$-brane at the region where $T\sim 0$.
The physical interpretation of this configuration is clear.
The kinky tachyon profile (\ref{kink}) corresponds to
a (non-BPS) $D8$-brane localized at $x^1\sim 0$.
\cite{Bergmank,Senv,Sen3,Sen4,Horava,Se} \
Since a fundamental string can end at the $D8$-brane, an
$(n,1)$ string can be changed to an $(n',1)$ string with
some integer $n'$ different from $n$
as the string crosses the $D8$-brane.
This $D8$-brane is unstable and it will disappear when the tachyon
on it condenses. In this case, the world-sheet theory on the $D1$-brane
probe describes a string junction as in \cite{Das}.
It is also possible to make a stable configuration with varying tachyon.
For example, the vortex configuration
\begin{eqnarray}
T=u(x^1+ix^2),~~~~~~(u\ra \infty)
\end{eqnarray}
for the $D9$-$\ol{D9}$ tachyon
is known to represent a BPS $D7$-brane.\cite{Senv,Wittenv,Se} \
The same argument as above can be applied to this configuration
when we place the $D1$-brane at $x^2=0$.

It is also interesting to consider a tachyon vortex parallel
to the $D1$-brane, such as
\begin{eqnarray}
T=u(x^8+i x^9),
\label{vortex89}
\end{eqnarray}
which makes a $D7$-brane localized at $x^8=x^9=0$.
When we place the $D1$-brane probe at $x^8=x^9=0$,
we obtain a massless Schwinger model as the
effective theory on the $D1$-brane world-sheet.
Actually, the $D1$-$D7$ system gives another realization of
the Schwinger model. The open string stretched between
the $D1$-brane and the $D7$-brane produces a Dirac fermion
$\lambda=(\lambda_+,\lambda_-)^T$
charged under the $U(1)$ gauge field on the $D1$-brane.
When we consider $n$ $D1$-branes on $N$ coincident $D7$-branes,
the massless fermions created by the $1$-$7$ strings
are as listed in table \ref{D7}.
\begin{table}[htb]
\begin{center}
$$
\begin{array}{c|cccc}
\hline\hline
\mbox{field}&U(n) & SO(1,1) & SO(2) & U(N)\\
\hline
\lambda_{+}&\textbf{fund.} & 1_+ & 1_+ &\textbf{fund.}\\
\lambda_{-}&\textbf{fund.} & 1_- & 1_- &\textbf{fund.}\\
\hline
\end{array}
$$
\parbox{75ex}{
\caption{
\small
The massless fields created by the $1$-$7$ string.
Here the $U(n)$ and $SO(1,1)$ are the gauge symmetry
and the Lorentz symmetry on the $D1$-brane, respectively.
The $SO(2)$ corresponds to the rotation of the
$x^8$-$x^9$ plane. The $U(N)$ is the gauge symmetry on
the $D7$-brane, which is seen as global symmetry
of the $D1$-brane.
}
\label{D7}
}
\end{center}
\end{table}
While the low energy field contents on the $D1$-brane world-sheet
are the same as those listed in table \ref{p=1},
the symmetry is different.
Here only the subgroup $SO(2)\times U(N)$ of the $U(N)\times U(N)$
chiral symmetry is manifest. The $SO(2)$ symmetry, which is the
rotational symmetry of the $x^8$-$x^9$ plane,
is now interpreted as the axial $U(1)_A$ symmetry
as we can see in the table \ref{D7}. This fact can also
be seen from the tachyon configuration (\ref{vortex89}).
The $U(1)_A$ acts on the tachyon field as a phase transformation
and it is translated to the rotation of the $x^8$-$x^9$ plane via
the relation (\ref{vortex89}).
Note that $D7$-brane charge is measured by the integration
of $dC_0$ along the $S^1$ surrounding the $D7$-brane as
\begin{eqnarray}
\int_{S^1} dC_0=N.
\end{eqnarray}
This implies that $C_0$ is shifted as
\begin{eqnarray}
C_0\ra C_0 + N\frac{\alpha}{\pi}
\label{C0shift}
\end{eqnarray}
under the rotation $x^8+ix^9 \ra e^{2i\alpha} (x^8+ix^9)$
which induces the $U(1)_A$ transformation (\ref{U1axial}).
This shift (\ref{C0shift}) precisely agrees with the shift
(\ref{shift}) for $N=1$ case and hence
the cancellation mechanism of the axial $U(1)_A$ anomaly
explained in section \ref{GS} also works in this case.

\subsection{Adding Flavors}
\label{multi}
It is not difficult to generalize our discussion to
the $N$ flavor case.
The bosonized action for the $N$ flavor (massive) Schwinger model is
given as \cite{Coleman,Abdalla}
\begin{eqnarray}
S=\int\! d^2 x \left\{\,
\half\sum_{I=1}^N\del_\mu\varphi_I\del^\mu\varphi_I
-\frac{m^2}{2}
\left(\sum_{I=1}^N \varphi_I\right)^2
+c\,T
 \left(\sum_{I=1}^N \cos\left(2\sqrt{\pi}\,\varphi_I
-\frac{\theta}{N}\right)-N\right)
\right\},\nn\\
\end{eqnarray}
where $m^{2}=\gym^{2}/\pi$ and we have assumed
that the tachyon field is proportional to the unit
matrix as
$T^I_{~\bar{I}}=T\,\delta^I_{~\bar I}$
with $T\in\R_{\ge 0}$.

When the tachyon condenses as $T\ra\infty$,
each scalar field $\varphi_I$ will take a discrete
value
\begin{eqnarray}
\varphi_I=\sqrt{\pi}\left(n_I+\frac{\theta}{2\pi N}\right),~~~(n_I\in\Z),
\end{eqnarray}
just as our discussion in the one flavor case (\ref{varphi}).
Then, we obtain the energy density carried by
this configuration as
\begin{equation}
{\cal E}=\frac{m^{2}}{2}\pi\left(\sum_{I=1}^N n_I+\frac{\theta}{2\pi}
\right)^2
=\frac{g_s}{4\pi\alpha'}\left(\sum_{I=1}^N n_I+C_0\right)^2,
\end{equation}
which is the same as the expression (\ref{ene})
with $n\equiv\sum_{I=1}^N n_I$. Therefore, this configuration
again represents an $(n,1)$ string. Actually,
the electric flux is related to the scalar fields as
\begin{equation}
F_{01}=-\frac{\gym^{2}}{\sqrt{\pi}}
\sum_{I=1}^N\varphi_I=-\gym^2(n+C_0),
\label{gaugef}
\end{equation}
which induces $n$ units of fundamental string charge as expected.

\section{QED${}_{2}$ on a Circle and
the T-dual Description \label{D0}}

In this subsection we compactify the direction
parallel to the $D1$-brane and give the string theory
interpretation of some of the old results in the Schwinger
model on $S^1$. We also consider applications
of the field theory results
to the $D0$-brane dynamics in the presence
of a $D8$-$\ol{D8}$ pair by taking T-duality.

\subsection{Schwinger Model on a Circle \label{circle}}
Here we recapitulate the prescription
given in \cite{Manton,Hosotani,Iso} for
the massless Schwinger model on a circle of radius $R$.
Since the spatial coordinate $x^1$ is now periodic as
$x^1\sim x^1+2\pi R$,
the spatial component of the gauge field $A_1$ also becomes a periodic
variable with period $1/R$ via the gauge transformation
\begin{eqnarray}
A_1\sim A_1+i\,e^{inx/R}(\del_1 e^{-inx/R})=A_1+\frac{n}{R}.
\label{periA}
\end{eqnarray}
Following \cite{Manton,Hosotani}, we choose a gauge with
\begin{eqnarray}
\del_1 A_1=0.
\label{gauge}
\end{eqnarray}
Note that we cannot make $A_1=0$ by a gauge transformation, since
$\exp\left(i\oint A_1 dx^1\right)$
is a gauge invariant quantity.
The equation of motion for $A_{0}$ is now
\begin{equation}
\del_1 F_{01}=-(\partial_{1})^{2}A_{0}=\gym^{2}\lambda^{\dagger}\lambda.
\label{eom}
\end{equation}
Then, standard manipulations lead to the Hamiltonian
\begin{eqnarray}
H=\int_0^{2\pi R}\! dx^1\left\{\, \frac{1}{2\gym^2}(\del_0 A_1)^2
-\lambda^{\dagger}i\sigma^{3}(\partial_{1}+iA_{1})\lambda
+\frac{\gym^2}{2}
(\lambda^\dag\lambda)\frac{1}{-\partial_{1}^{2}}
 (\lambda^\dag\lambda)\,\right\}.
\label{Hami}
\end{eqnarray}
Here we have eliminated $A_0$ in the Hamiltonian using (\ref{eom}).

It is also useful to work in the momentum representation.
The Fourier expansion of fermion fields at a fixed time slice,
say $x^0=0$, is given by
\begin{eqnarray}
\lambda(x)=\frac{1}{\sqrt{2\pi R}}
\sum_{k\in\Z}
a_k\, e^{i\frac{k}{R}x^1},
\label{operator}
\end{eqnarray}
where $a_k=(a_{+,k},a_{-,k})^T$ satisfy the canonical anti-commutation
relations
\begin{eqnarray}
\{a_{\alpha,k}^{\dagger},a_{\beta,l}\}=\delta_{kl}\delta_{\alpha\beta},
~~~
\{a_{\alpha,k},a_{\beta,l}\}=\{a_{\alpha,k}^{\dagger},
a_{\beta,l}^{\dagger}\}=0.
\label{CAR}
\end{eqnarray}
Then, the Hamiltonian (\ref{Hami}) is written as
\begin{eqnarray}
H=\frac{2\pi R}{2\gym^2}(\del_0 A_1)^2
+\sum_{k \in\Z}\left(\frac{k}{R}+A_{1}\right)
a_k^\dag \sigma^3 a_k
+\frac{\gym^{2}}{2}\sum_{k \neq 0}
j^0(k)\frac{R^{2}}{k^{2}}j^0(-k).
\label{momHam}
\end{eqnarray}
Note that $\del_0 A_1$ is the zero-momentum component of
the electric field $F_{01}$ in our gauge (\ref{gauge}).
{}From (\ref{momHam}), we see that 
the operators $a_{\pm,k}^{\dag}$ and  $a_{\pm,k}$ are creation and 
annihilation operators for a positive/negative chirality particle of 
momentum $k/R$ and energy $\pm(k/R+A_{1})$, respectively.

The bosonization techniques can also be applied to this system. 
The relations (\ref{phipi}) and  (\ref{currentphi}) in the momentum
representation are 
\begin{eqnarray}
\varphi(k)&=&
\left\{
\begin{array}{cc}
\frac{1}{\sqrt{2R}}\frac{2\pi R}{\gym^2}\,\del_0A_1& (k=0)\\
\vspace{-3mm}\\
-i\sqrt{\pi}\frac{R}{k}\, j^0(k)& (k\ne 0)
\end{array}
\right.\\
\pi_\varphi(k)&=&\sqrt{\pi}\, j_5^0(k).
\end{eqnarray}
Then,
it is easy to see that the first and the third terms in the Hamiltonian
(\ref{momHam})
correspond to the $k=0$ and $k\ne 0$ parts of the mass term of the scalar
field $\varphi$, respectively, and we have
\begin{eqnarray}
\frac{2\pi R}{2\gym^2}(\del_0 A_1)^2+
\frac{\gym^{2}}{2}\sum_{k \neq 0}j^0(k)\frac{R^{2}}{k^2}j^0(-k)
=\frac{\gym^2}{2\pi} \sum_{k\in\Z}\varphi(k)\varphi(-k).
\label{1st3rd}
\end{eqnarray}
The second term in the Hamiltonian (\ref{momHam}) is more involved
and one should carefully regularize the infinite sum.
Here we just present the result shown in \cite{Manton};
\begin{eqnarray}
\sum_{k\in\Z}\left(\frac{k}{R}+A_1\right) a_k^\dag\sigma^3 a_k
=\half\sum_{k\in\Z}\left(\pi_\varphi^\dag(k)\pi_\varphi(k)
+\left(\frac{k}{R}\right)^2\varphi^\dag(k)\varphi(k)\right),
\label{2nd}
\end{eqnarray}
neglecting an additive constant term.
The equations (\ref{1st3rd}) and (\ref{2nd})
show that the Hamiltonian (\ref{momHam}) is equivalent to that
for the free scalar field
\begin{eqnarray}
H=\half\sum_{k\in\Z}
\left(
\pi_\varphi^\dag(k)\pi_\varphi(k)
+\left(\frac{k}{R}\right)^2\varphi^\dag(k)\varphi(k)
+\frac{\gym^2}{\pi}\varphi^\dag(k)\varphi(k)
\right).
\label{phiHam2}
\end{eqnarray}

\subsection{Fermion Fock Space and the Hanany-Witten Effect
\label{Fermion}}

The fermion Fock space for the Schwinger model on $S^1$
is constructed by acting the operators
 $a_{\pm,k}^{\dag}$ and  $a_{\pm,k}$ 
 satisfying (\ref{CAR})
on a Fock vacuum defined by
\begin{eqnarray}
a_{+,k}\ket{M,N;A_1}&=&0~~~~(\mbox{for}~~k>M),\nn\\
a^\dag_{+,k}\ket{M,N;A_1}&=&0~~~~(\mbox{for}~~k\le M),\nn\\
a_{-,k}\ket{M,N;A_1}&=&0~~~~(\mbox{for}~~k<N),\nn\\
a^\dag_{-,k}\ket{M,N;A_1}&=&0~~~~(\mbox{for}~~k\ge N)
\end{eqnarray}
for a fixed value of $A_1\in [0,1/R]$.
 By definition, these states satisfy
\begin{eqnarray}
\ket{M,N;A_{1}}= a_{+,M}^{\dagger}\ket{M-1,N;A_{1}}
=a_{-,N-1}\ket{M,N-1;A_{1}}.
\label{shift10}
\end{eqnarray}
In addition, since the gauge transformation with $g=e^{ix/R}$ induces
$A_1\ra A_1+1/R$ and $a_{\pm,k}\ra a_{\pm,k+1}$,
the Fock vacua satisfy the boundary condition
\begin{eqnarray}
\ket{M,N;0}=\ket{M-1,N-1;1/R}.
\label{bdry}
\end{eqnarray}

The total electric charge and axial charge is given by 
\begin{eqnarray}
Q=Q_++Q_-, ~~~
Q_{5}=Q_+-Q_-,
\end{eqnarray}
respectively, where
\begin{equation}
Q_\pm\equiv\int_0^{2\pi R}\! dx^1\, \lambda_\pm^\dag(x)\lambda_\pm(x)
=\sum_{k\in\Z}a_{\pm,k}^{\dag}a_{\pm,k}.
\end{equation}
With an appropriate regularization, it was shown in \cite{Manton} that
\begin{eqnarray}
Q_+\ket{M,N;A_1}&=&\left(M+RA_1+\half\right)\ket{M,N;A_1}\\
Q_-\ket{M,N;A_1}&=&\left(-N-RA_1+\half\right)\ket{M,N;A_1}.
\end{eqnarray}
Since the equation (\ref{eom}) implies $Q=0$, we take $N=M+1$
and define a state
\begin{eqnarray}
\ket{\wt A}=\ket{M,M+1;A_1},
\end{eqnarray}
where $\wt A=A_1+\frac{M}{R}$. Note that the right hand side
depends only on $\wt A$ because of the condition (\ref{bdry})
and hence the state $\ket{\wt A}$ is well-defined.
{}From (\ref{shift10}), it satisfies
\begin{eqnarray}
\ket{\wt A+1}=a^\dag_{+,M+1}a_{-,M+1}\ket{\wt A}.
\label{shift2}
\end{eqnarray}
The axial charge of this state is given by
\begin{eqnarray}
Q_5\ket{\wt A}=2\left(R\wt A+\half\right)\ket{\wt A}.
\label{q5}
\end{eqnarray}

Then, the Hamiltonian for the wave function
$\psi(\wt A)=\Bracket{\wt A}{\psi}$ becomes \cite{Manton}
\begin{eqnarray}
\VEV{\wt A\,\Big|\,H\,\Big|\, \psi}=
\left(-\frac{\gym^2}{4\pi R}
\frac{\del^2}{\del {\wt A}^2}+V(\wt A)\right)\psi(\wt A),
\label{HamiPsi}
\end{eqnarray}
where
\begin{eqnarray}
V(\wt A)=\frac{1}{R}\left(
R\wt A+\half\right)^2.
\label{pot}
\end{eqnarray}
Here, only the zero-momentum part of the Hamiltonian (\ref{phiHam2})
contributes and we have used the relations
\begin{eqnarray}
&&\varphi(k=0)=\frac{1}{\sqrt{2R}}\,
\frac{2\pi R}{\gym^2}\del_0 A_1=\frac{-i}{\sqrt{2R}}\,
\frac{\del}{\del\wt A},\\
&&\pi_\varphi(k=0)=\frac{1}{\sqrt{2R}}\,Q_5
=\sqrt{\frac{2}{R}}\left(R\wt A+\half\right)
\end{eqnarray}
as the operators acting on $\ket{\wt A}$.

It would be more convenient to express these in
the T-dual picture. When we T-dualize the compactified
$x^1$-direction, we obtain a system with a $D0$-brane
and a $D8$-$\ol{D8}$ pair, which are localized on a circle of
radius $\wh R\equiv\alpha'/R$.
The gauge field $A_1$ on the $D1$-brane
is related to the position $X^1$ of the $D0$-brane as
$A^1=\frac{X^1}{2\pi\alpha'}$.\cite{PoChJo} (See also
section \ref{T-duality})
The gauge transformation (\ref{periA}) implies
the periodicity of the T-dualized circle $X^1\sim X^1+2\pi\wh R$.
We also introduce the notation
$\wt X\equiv 2\pi\alpha'\wt A$
which corresponds to
the position of the $D0$-brane in the covering space
of the $S^1$.
Then the potential energy (\ref{pot}) of the system is rewritten as
\begin{eqnarray}
V(\wt X)=\frac{T_1}{2\pi\wh R}\left(
\wt X+\pi\wh R \right)^2.
\label{potX}
\end{eqnarray}
The momentum index $k$ carried by
the operators $a_{\pm,k}$ and $a^\dag_{\pm,k}$
is mapped to the winding number and
the energy $\pm(k/R+A_1)=\pm T_1(2\pi k\wh R+X^1)$ is interpreted as
the length times the tension $T_1\equiv 1/2\pi\alpha'$
of the $0$-$8$ or $0$-$\ol 8$ strings.
\begin{table}[htb]
\begin{center}
$$
\begin{array}{ccccc}
\hline\hline
\mbox{operator}&\mbox{energy}&\mbox{winding}&\mbox{charge}&\mbox{string}\\
\hline
\vspace{-4mm}\\
a_{+,k}^\dag&T_1(2\pi k\wh{R}+X^{1})&\left(k+\frac{X^{1}}{2\pi\hat R}\right)&
+&\mbox{$0$-$8$ string}  \\
\vspace{-4mm}\\
a_{+,k}     &-T_1(2\pi k\wh{R}+X^{1})&-\left(k+\frac{X^{1}}{2\pi\hat R}\right)&
-&\mbox{$8$-$0$ string}    \\
\vspace{-4mm}\\
a_{-,k}^\dag&-T_1(2\pi k\wh{R}+X^{1})&\left(k+\frac{X^{1}}{2\pi\hat R}\right)&
+&\mbox{$0$-$\ol 8$ string}    \\
\vspace{-4mm}\\
a_{-,k}     &T_1(2\pi k\wh{R}+X^{1})&-\left(k+\frac{X^{1}}{2\pi\hat R}\right)&
-&\mbox{$\ol 8$-$0$ string}    \\
\vspace{-4mm}\\
\hline
\end{array}
$$
\parbox{75ex}{
\caption{\small
The fermionic operators in (\ref{CAR}) carry
 the energy, winding number, and charge as in this table.
We can identify the states created
 by these operators as the states generated by the
 corresponding oriented strings
 stretched between $D0$ and $D8$
 or $D0$ and $\ol{D8}$ as in this table.
 }
\label{table3}}
\end{center}
\end{table}

\begin{figure}[htb]
\begin{center}
\parbox{3cm}{
\unitlength=.4mm
\begin{picture}(50,75)(0,0)
\epsfxsize=2cm
\put(0,15){\epsfbox{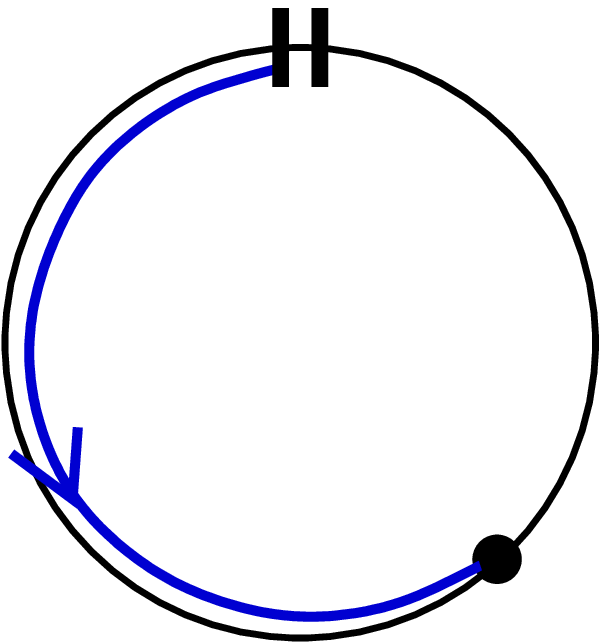}}
\put(52,17){\makebox(0,0){$D0$}}
\put(25,75){\makebox(0,0){$D8$-$\ol{D8}$}}
\put(24,0){\makebox(0,0){
\footnotesize $a^\dag_{+,0}\ket{\wt X}$}}
\end{picture}
}
\parbox{3cm}{
\unitlength=.4mm
\begin{picture}(50,75)(0,0)
\epsfxsize=2cm
\put(0,15){\epsfbox{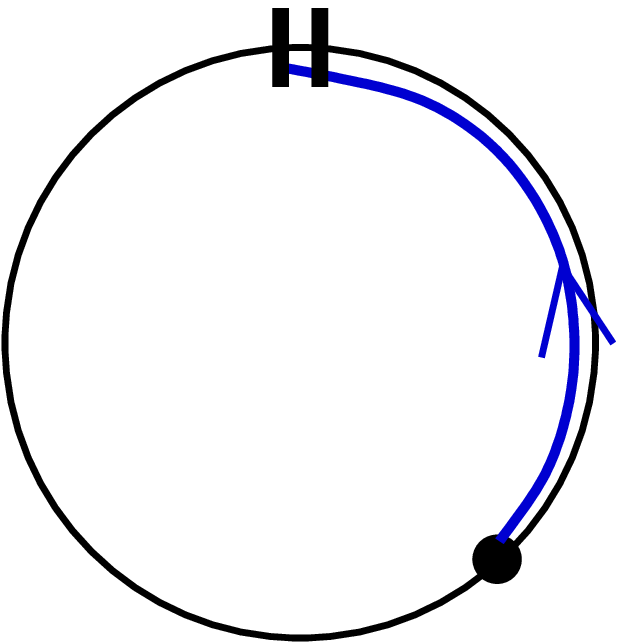}}
\put(52,17){\makebox(0,0){$D0$}}
\put(25,75){\makebox(0,0){$D8$-$\ol{D8}$}}
\put(24,0){\makebox(0,0){
\footnotesize $a_{+,-1}\ket{\wt X}$}}
\end{picture}
}
\parbox{3cm}{
\unitlength=.4mm
\begin{picture}(50,75)(0,0)
\epsfxsize=2cm
\put(0,15){\epsfbox{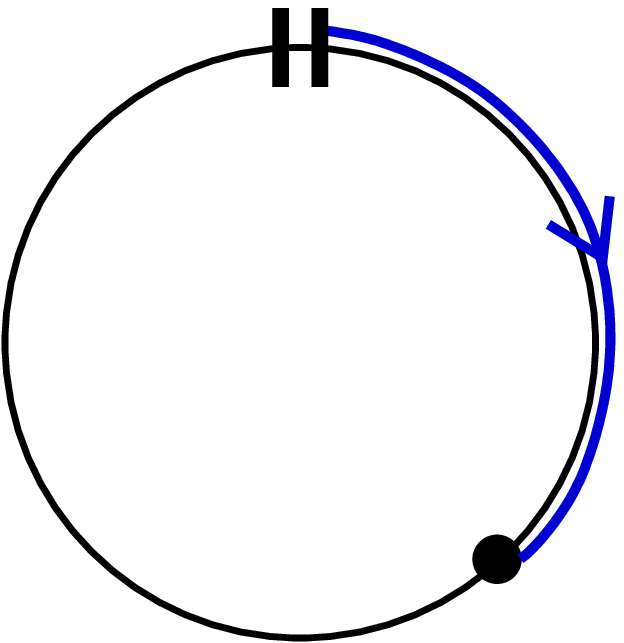}}
\put(52,17){\makebox(0,0){$D0$}}
\put(25,75){\makebox(0,0){$D8$-$\ol{D8}$}}
\put(24,0){\makebox(0,0){
\footnotesize $a^\dag_{-,-1}\ket{\wt X}$}}
\end{picture}
}
\parbox{3cm}{
\unitlength=.4mm
\begin{picture}(50,75)(0,0)
\epsfxsize=2cm
\put(0,15){\epsfbox{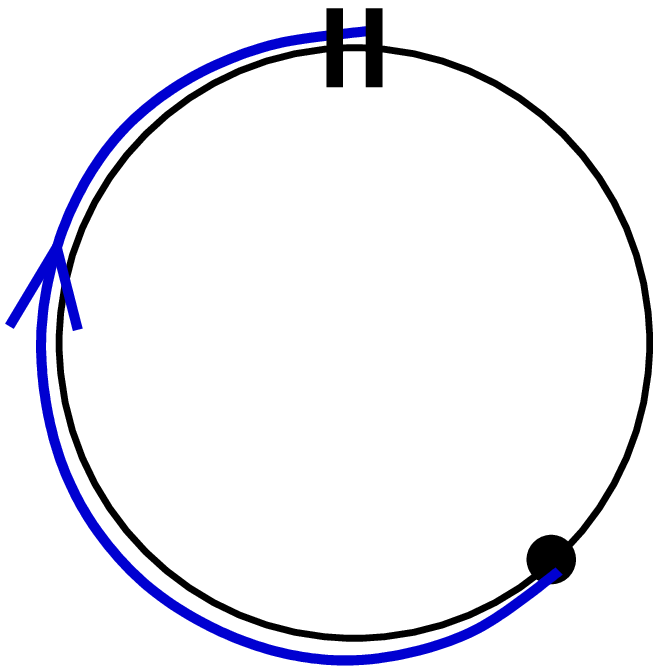}}
\put(52,17){\makebox(0,0){$D0$}}
\put(25,75){\makebox(0,0){$D8$-$\ol{D8}$}}
\put(24,0){\makebox(0,0){
\footnotesize $a_{-,0}\ket{\wt X}$}}
\end{picture}
}
\end{center}
\begin{center}
\parbox{75ex}{
\caption{\small
{}From table \ref{table3}, we can identify the fermionic states
 created by the operators in (\ref{CAR})
 as the states generated by the corresponding strings
 as in this figure.
 }\label{fig1}}
\end{center}
\end{figure}

In the classical picture, the excited
energy of the system with a static $D0$-brane is expected to
be proportional to the length of the strings attached on it.
Since the minimum of the potential (\ref{potX}) is given by
 $\wt X=-\pi\wh R$, it is natural to interpret that the state
$\ket{\wt X}\equiv\ket{\wt A}$ with $-2\pi\wh R<\wt X<0$
corresponds to a configuration without any strings attached
on the $D0$-brane.
The first excited states are
\begin{eqnarray}
a^\dag_{+,0}\ket{\wt X},~~~
a_{+,-1}\ket{\wt X},~~~
a^\dag_{-,-1}\ket{\wt X},~~~
a_{-,0}\ket{\wt X},
\end{eqnarray}
which correspond to the configurations with one string
attached on the $D0$-brane as depicted in Figure \ref{fig1}.
Note however that these states are not physical since the total electric
charge does not vanish.

What happens when the $D0$-brane turns around the circle\,?
As we have seen in (\ref{shift2}), $2\pi\wh R$ shift of the
position of the $D0$-brane can be expressed as
\begin{eqnarray}
\ket{\wt X+2\pi\wh R}=a_{+,0}^\dag a_{-,0}\ket{\wt X}
\end{eqnarray}
for $-2\pi\wh R<\wt X<0$. The right hand side of this equation
corresponds to the configuration with a $0$-$8$ string and
a $\ol 8$-$0$ string attached on the $D0$-brane.
This relation can be understood as
a result of the Hanany-Witten effect.\cite{Hanany-Witten}
\footnote{
See also \cite{Bachas2,Dani,Bergman,HoWu,Ohta1,Bergman2,Bachas,Ohta2}
for the discussions
in closely related situations.}
It suggests that
when the $D0$-brane crosses the $D8$-brane or $\ol{D8}$-brane,
a fundamental string is created between the $D0$-brane
and the $D8$-brane or $\ol{D8}$-brane. Actually,
the axial charge is shifted by two when the $D0$-brane
travels around the circle due to the axial $U(1)_A$ anomaly
as we can explicitly see from (\ref{q5}) which implies
the creation of a pair of fundamental strings
(a $0$-$8$ string and a $\ol 8$-$0$ string)
as discussed in \cite{Bachas2}.
It is also interesting to note that we cannot make a configuration
with two or more strings of the same type attached on the $D0$-brane
because of the Pauli's exclusive principle, since the operators
$a_{\pm,k}$ are fermionic. This observation is used as the explanation of
the s-rule in \cite{Bachas,Bachas10},
though our configuration is non-supersymmetric.

When the $D0$-brane is turned $k$ times around
the circle, the system will gain the energy of
\begin{eqnarray}
V(2\pi k\wh R+\wt X)-V(\wt X)&=&2T_1 k\left(\wt X+(k+1)\pi\wh R\right)\\
&=&2T_1\sum_{l=1}^k \left(2\pi l\wh R+\wt X\right).
\label{HaWiPot}
\end{eqnarray}
Note that the length of the strings created by the Hanany-Witten effect
during the process is $(2\pi l\wh R+\wt X)$ ($l=1,2,\dots,k$).
The expression (\ref{HaWiPot}) clearly shows that it is equal to
the energy carried by these strings.

In order to obtain the precise form of the potential (\ref{potX}),
we have to take into account the quantum effects.
A string theoretical derivation of the potential will be given in
the next subsection.

\subsection{The $D0$-brane Potential from Stringy Calculations
 \label{S-cal}}

Let us suppose $\wt X=2\pi k\wh R+ X$ with $k\in\Z$, $-2\pi\wh R<X<0$ and
extract the quantum correction of the potential
\begin{eqnarray}
\Delta V(\wt X)& \equiv &V(\wt X)-2T_1\sum_{l=1}^k\left(2\pi l\wh R+X\right)
=\frac{T_1}{2\pi\wh R}\left(X+\pi\wh R \right)^2\\
&=&\frac{2\wh RT_1}{\pi}\sum_{n=1}^\infty\frac{1}{n^2}\cos\left(
\frac{n\wt X}{\wh R}
\right)+ \mbox{const.},
\label{D0pot}\\
&=&\frac{1}{\pi^2 R}\sum_{n=1}^\infty\frac{1}{n^2}\cos\left(
2\pi n R\wt A
\right)+ \mbox{const.},
\label{Qpot}
\end{eqnarray}
by subtracting the energy carried by the fundamental strings
 (\ref{HaWiPot})
from the potential $V(\wt X)$ (\ref{potX}).
The expression (\ref{Qpot}) is obtained in \cite{Hosotani}
{}from the one-loop diagrams of fermions in the
Schwinger model on $S^1$. 
It would be instructive to 
rederive the $D0$-brane potential (\ref{D0pot})
in terms of string theory.

The potential of the $D0$-brane in the presence of 
the $D8$-$\ol{D8}$ pair is produced by
the exchange of closed strings represented by 
the amplitude of the cylinder diagram
\begin{equation}
{\cal A}(\wt{X})=\int_0^{\infty}d \tau
 \left(\bra{D0} e^{-\pi \tau (L_{0}
+\wt{L_{0}})} \ket{D8}
+\bra{D0} e^{-\pi \tau (L_0+\wt{L_0})}
 \ket{\ol{D8}}\right),
\end{equation}
where $\ket{D0}$, $\ket{D8}$ and $\ket{\ol{D8}}$
are the boundary states corresponding to the $D0$-brane,
$D8$-brane and $\ol{D8}$-brane, respectively.
This amplitude for the non-compact case
has been calculated in \cite{Bergman} and
the result is
\begin{eqnarray}
{\cal A}_{\mbox{\tiny non-cpt}}(X)=-\frac{V_0}{(8\pi^2\alpha')^{1/2}}\int_0^\infty
d\tau\, \tau^{-1/2}e^{-X^2/2\pi\alpha'\tau},
\end{eqnarray}
where $X$ is the distance between the $D0$-brane
and the $D8$-$\ol{D8}$ pair and $V_0$ is the volume factor.
 Note that we haven't taken
$\alpha\ra 0$ limit here. The contribution
of the massive closed strings automatically cancels
in this cylinder amplitude.

When we compactify the direction transverse to the branes
as in the previous subsection, we have to take into account
the contributions from the infinite copies of the
$D8$-$\ol{D8}$ pairs in the covering space as
\begin{eqnarray}
{\cal A}(\wt X)=\sum_{n\in\Z}
{\cal A}_{\mbox{\tiny non-cpt}}(\wt X+2\pi\wh R n).
\end{eqnarray}
Then the potential will become
\begin{eqnarray}
\Delta V(\wt X)&=&-{\cal A}(\wt X)/V_0\\
&=&\frac{1}{(8\pi\alpha')^{1/2}}\sum_{n\in\Z}
\int_0^\infty\!d\tau\,
\tau^{-1/2} e^{-(\tilde X+2\pi\hat Rn)^2/2\pi\alpha'\tau}
\label{poisson1}\\
&=&\frac{1}{(8\pi\alpha')^{1/2}}\frac{\sqrt{2\alpha'}}{2\wh R}
\sum_{n\in\Z}\int_0^\infty \!d\tau \,
e^{\frac{\pi\alpha' n^2}{2\hat R^2}\tau-\frac{n\tilde X}{R}i}
\label{poisson2}\\
&=&\frac{2\wh R}{\pi (2\pi\alpha')}\sum_{n=1}^\infty
\frac{1}{n^2}\cos\left(\frac{n\wt X}{\wh R}\right)+\mbox{const.},
\label{D0pot2}
\end{eqnarray}
reproducing (\ref{D0pot}) as promised.
Here we have used the Poisson resummation formula in
the equality between (\ref{poisson1}) and (\ref{poisson2}).

\subsection{Quantum Mechanics of the $D0$-brane
 in the $D8$-$\overline{D8}$ Background \label{QM}}

Here we continue our discussion in the T-dual picture.
The wave function $\psi(\wt X)=\Bracket{\wt X}{\psi}$
is now regarded as the wave function of the $D0$-brane.
T-dualizing the Hamiltonian (\ref{HamiPsi}), we obtain
the Schr\"{o}dinger equation for the $D0$-brane wave function as
\begin{eqnarray}
i \frac{\partial}{\partial t} \psi(\wt{X})
=\left(-\frac{1}{2T_{0}}\frac{\del^2}{\del \wt X^2}
+V(\wt{X})\right) \psi(\wt{X}),
\end{eqnarray}
where $T_0\equiv 1/\wh g_s l_s$ ~($\wh{g}_s\equiv g_s l_s/R$)
is the mass of the $D0$-brane
as $\wh{g}_s$ is the string coupling in the T-dual picture.
Since the potential (\ref{potX}) is quadratic with respect to 
$\wt X+\pi \wh R$, this Schr\"odinger equation is the same as that
for a harmonic oscillator.
The ground state is given by
\begin{eqnarray}
\psi(\wt X)&=&{\cal N}
\exp\left\{
-\frac{(\wt X+\pi\wh R)^2}{\sqrt{8\pi^2\wh{g}_sl_s^3\wh R}}
\right\}
\label{D0wave}\\
&=&{\cal N}
\exp\left\{-\frac{\sqrt{\pi}R}{\gym}
\left(\wt A+\frac{1}{2R}\right)^2
\right\}
\end{eqnarray}
where $\cal N$ is a normalization constant.

According to (\ref{D0wave}), the $D0$-brane is trapped around
$\wt X=-\pi\wh R$ with a width of order
$(\wh{g}_sl_s^3\wh R)^{1/4} \sim \alpha'(\gym/R)^{1/2}$.
Though this width of the quantum fluctuation will formally
become zero if we take the decoupling limit $\alpha'\ra 0$
keeping $\gym$ and $R$ fixed, the ratio to the radius $\wh R=\alpha'/R$
remains finite as $\delta \wt X/\wh R\sim (R\gym)^{1/2}$.

\section{T-duality,
Matrix and Lattice Regularization
\label{Matrix}}

One of the advantages of the realization of a field theory
in string theory is that we can apply various dualities
known in string theory.
As an application of our realization of QED on a D-brane
world-volume, we can apply T-duality and obtain
its matrix theory description. Actually, this method can be
applied to a wide variety of field theories including
the realistic QCD, even though we haven't obtained its
string theory realization.\footnote{See \cite{Evans,Kruczenski:2003uq}
for recent progress.}
We also show that when we regularize the theory by replacing
infinite size matrices to finite ones, we naturally
obtain usual lattice gauge theory.

Although we use some terminology appeared before
to make connections with the previous sections,
most part of this section can be read independently.
Actually, this section is intended to be readable
for the readers who may not be familiar with string theory.

Throughout this section we have set $2\pi\alpha'=1$.

\subsection{T-duality and Matrix Theory Description of QED
\label{T-duality}}

Let us first consider 4 dimensional QED with $N_f$ flavors as an exercise,
whose action is
\begin{equation}
S_{D3}=\int{d^{4}x}\left(-\frac{1}{4\gym^{2}}
 F_{\mu\nu}F^{\mu\nu}
+i\ol{\lambda}_I\gamma^{\mu}D_\mu\lambda^I\right),
\label{4Daction}
\end{equation}
where $D_\mu=\del_\mu+iA_\mu$.
As discussed in section \ref{twoqed}, it is realized as
the world-volume theory of a $D3$-brane in the background of
$N_f$ $D9$-$\ol{D9}$ pairs. When we compactify one of the spatial
directions, say $x^3$, to $S^1$ of radius $R$
and take T-duality along it, we obtain a system
with a $D2$-brane and $N_f$ $D8$-$\ol{D8}$ pairs localized on the $S^1$
of radius $\wh R=1/2\pi R$
and the 4 dimensional QED is mapped to a 3 dimensional theory
realized on the $D2$-brane. Then, we can apply the
prescription given in \cite{Taylor1,Banks,Ganor,Taylor2}
to obtain this map. Let us explain this prescription shortly.
The 3 dimensional world-volume theory consists of the following
infinite size matrix valued fields
\begin{eqnarray}
(A_\alpha(x^\beta))^A_{~B},~~~(X^3(x^\beta))^A_{~B},
~~~(\lambda^I(x^\beta))^A_{~B},
~~~~~(\alpha,\beta=0,1,2;~A,B\in\Z),
\label{AXL}
\end{eqnarray}
satisfying the constraints
\begin{eqnarray}
(A_\alpha(x^\beta))^{A+1}_{~~B+1}&=&(A_\alpha(x^\beta))^{A}_{~B},\\
(X^3(x^\beta))^{A+1}_{~~B+1}&=&(X^3(x^\beta))^{A}_{~B}
+2\pi\wh R\,\delta^A_{~B},\label{Xcon}\\
(\lambda^I(x^\beta))^{A+1}_{~~B+1}&=&(\lambda^I(x^\beta))^{A}_{~B}.
\end{eqnarray}
A trivial solution of the constraint (\ref{Xcon}) is given by
a diagonal matrix  $(\Delta^3)^A_{~B}=2\pi\wh R A\, \delta^A_{~B}$.
Using this matrix, the constraint (\ref{Xcon}) can also be written as
\begin{eqnarray}
(\wh X^3(x^\beta))^{A+1}_{~~B+1}=(\wh X^3(x^\beta))^{A}_{~B},~~~~~
\wh X^3(x^\beta)\equiv X^3(x^\beta)-\Delta^3.
\end{eqnarray}
{}From these constraints, we can choose a row (e.g. $B=0$)
of each matrix in (\ref{AXL})
as the independent degrees of freedom.

The explicit correspondence between the 4 dimensional fields
and the 3 dimensional fields
is given by the Fourier transformation as
\begin{eqnarray}
A_\alpha(x^\beta,x^3)&=&\sum_{A\in\Z}
(A_\alpha(x^\beta))^A_{~0}\,e^{iA x^3/R}, \label{F1}\\
A_3(x^\beta,x^3)&=&
\sum_{A\in\Z} (X_3(x^\beta))^A_{~0}\,e^{iA x^3/R},\\
\lambda^I(x^\beta,x^3)
&=&\sum_{A\in\Z} (\lambda^I(x^\beta))^A_{~0}\,e^{iA x^3/R}.
\label{F3}
\end{eqnarray}
Note that the gauge choice $\del_3A_3=0$ we used in (\ref{gauge})
for the Schwinger model implies $X^3$ is diagonal and
$A_3=(X^3)^0_{~0}$ which is the relation
we used in section \ref{T-duality}.
The action for the 3 dimensional description is
\begin{eqnarray}
S_{D2}&=&\frac{2\pi R}{\Tr 1}
\int{d^{3}x}\,\Tr\left(-\frac{1}{4\gym^{2}}
 F_{\alpha\beta}F^{\alpha\beta}
+i\ol{\lambda}_I\gamma^{\alpha}D_\alpha\lambda^I
\right.\nn\\
&&~~~~~~~\left.
-\frac{1}{4\gym^2}[D_\alpha, X_3][D^\alpha, X^3]
+i\ol{\lambda}_I\gamma^{3}\wh D_3\lambda^I
\frac{}{}\right),
\label{3Daction}
\end{eqnarray}
where
$\wh D_3\lambda^I \equiv i([\Delta^3,\lambda^I]+\wh X^3\lambda^I)$.
Here we need to divide the trace by $\Tr 1$ to extract one component
out of the infinite copies in the covering space.
One can easily show that the actions (\ref{4Daction}) and (\ref{3Daction})
are equal under the correspondence (\ref{F1})$\sim$(\ref{F3}).

We can T-dualize the rest of the directions in the same way.
When we take T-duality along all the space-time directions,
\footnote{To perform T-duality along the time direction,
we consider a Wick rotated theory and formally apply the T-duality
rules to the Euclidean time direction.}
we obtain a matrix theory description of the 4 dimensional
QED based on the $D(-1)$-branes in the presence of
$D5$-$\ol{D5}$ pairs.
Then, the world-volume is 0 dimensional and the theory is described
by infinite size matrices $(X^\mu)^A_{~B}$ ($\mu=0,\dots,3$)
and $(\lambda^I)^A_{~B}$
constrained as
\begin{eqnarray}
(X^{\mu})^{A+\hat{\nu}}_{\quad B+\hat{\nu}}&=&
(X^{\mu})^{A}_{~B}+2 \pi \wh{R}\,\delta^\mu_{~\nu}\delta^A_{~B},
\label{con1}\\
(\lambda^I)^{A+\hat{\nu}}_{\quad B+\hat{\nu}}&=&
(\lambda^I)^{A}_{~B}.
\label{con2}
\end{eqnarray}
Here the indices $A,B$ are labeled by $\Z^4$ as
$A=(a_{0},a_{1},a_{2},a_{3})$ with $a_{0},a_{1},a_{2},a_{3}
\in \Z$ and we have used the notation $\wh\nu=
(\delta_{\nu 0},\delta_{\nu 1},\delta_{\nu 2},\delta_{\nu 3})$.

The action is now
\begin{equation}
S_{D(-1)}= \frac{(2\pi R)^{4}}{\Tr 1}\Tr \left(\frac{1}{4\gym^{2}}
[X_{\mu},X_{\nu}][X^{\mu},X^{\nu}]
 +i\ol\lambda_I\gamma^{\mu}\wh D_{\mu}\lambda^I \right).
\label{matrix}
\end{equation}
where the covariant derivative of $\lambda^I$ is given as
\begin{eqnarray}
\wh D_\mu\lambda^I& \equiv&i([\Delta_\mu,\lambda^I]+\wh X_\mu\lambda^I)
=i(X_\mu\lambda^I-\lambda^I\Delta_\mu),\\
(\Delta_\mu)^A_{~B}&\equiv&2\pi\wh R\, a_\mu\delta^A_{~B}.
\label{Delta}
\end{eqnarray}

We have explained the T-duality prescription using 4 dimensional QED.
The generalization to arbitrary gauge theory is straightforward
since it is just a Fourier transformation.

\subsection{Matrix Regularization and Lattice Gauge Theory
\label{orbifold}}

The matrix theory description of the gauge field theory (\ref{matrix})
is not practical for computer simulations, since the size of
matrices is infinite.
Here we consider a natural regularization to cut off the
size of the matrices.

We again use the 4 dimensional QED considered in the previous
subsection as an illustrative example,
though the generalization to other field theory is straightforward.
The basic idea is to replace the group $\Z^4$ which labels
the matrix indices with a finite group $\Z_N^4$. Then,
the problem is that the constraint (\ref{con1}) does not
make sense if the indices $A,B$ were labeled by $\Z_N^4$.
Actually, it implies
\begin{eqnarray}
(X^{\mu})^{A+N\hat{\mu}}_{\quad B+N\hat{\mu}}&=&
(X^{\mu})^{A}_{~B}+2 \pi \wh{R}N\,\delta^A_{~B}.
\end{eqnarray}
While the left hand side should be equal to $(X^{\mu})^{A}_{~B}$
when the indices $A,B$ are $\Z_N^4$ valued,
the second term of the right hand side $2\pi\wh RN$ cannot be zero.
A possible modification is to exponentiate the matrices $X^\mu$
and introduce $U(N)$ matrices
\begin{eqnarray}
U_\mu\equiv e^{iaX^\mu},~~~~ a\equiv \frac{1}{\wh RN}
\end{eqnarray}
which are subject to the constraints
\begin{eqnarray}
(U_{\mu})^{A+\hat{\mu}}_{\quad B+\hat{\mu}}&=&
 \omega\, (U_{\mu})^{A}_{\; B}, \nonumber \\
(U_{\mu})^{A+\hat{\nu}}_{\quad B+\hat{\nu}}&=&
(U_{\mu})^{A}_{\; B},
\quad \mbox{for} \quad \mu \neq \nu,
\label{Uconst}
\end{eqnarray}
where $\omega =e^{2 \pi i / N}$. Here $a=1/\wh RN$ is a small parameter
which becomes zero in the large $N$ limit. We will soon
see that this parameter $a$ is interpreted as the lattice spacing.
Note that $aN=2\pi R$ is the length of the space of the 4 dimensional
theory, which is fixed in the large $N$ limit.

We can construct a (naive) action which reproduce (\ref{matrix}) at
leading order in $a$ as
\begin{equation}
S= a^4\left(\frac{1}{4\gym^2 a^4}
\sum_{\mu\ne\nu}\Tr\left(U_{\mu\nu}+U_{\mu\nu}^\dag-2\right)
 +\frac{1}{2ia}\Tr
 \left(\ol{\lambda}_I\gamma^{\mu}
(U_{\mu}\lambda^I\Omega_\mu^{-1}
-U_{\mu}^{\dag}\lambda^I\Omega_\mu)\,
\right)\right),
\label{Uaction}
\end{equation}
where $U_{\mu\nu}=U_\mu U_\nu U_\mu^\dag U_\nu^\dag$
and $\Omega_\mu$ is defined as
\begin{eqnarray}
(\Omega_\mu)^A_{~B} =(e^{ia\Delta_\mu})^A_{~B}=\omega^{a_\mu}\delta^A_{~B}.
\label{Ome}
\end{eqnarray}
This action is closely related to the
unitary matrix model proposed in \cite{Poly,Kit,Tada}.
Just like the usual
lattice theory, this action (\ref{Uaction})
suffers from the fermion doubling problem.
It could be avoided in a similar way as in the usual
lattice theory, say adding a Wilson term,
though we will not discuss it here.
This action also reminds us of the Eguchi-Kawai model \cite{EK},
since the action (\ref{Uaction}) looks like
that obtained
as the dimensional reduction of 4 dimensional
$U(N)$ gauge theory. But, here we have the constraints
(\ref{con2}) and (\ref{Uconst}) for the matrices.

Now we claim that the action (\ref{Uaction}) together with
the constraints (\ref{con2}) and (\ref{Uconst}) defines the
usual lattice regularization of the action (\ref{4Daction}).
Actually, the matrix description (\ref{Uaction})
 is the momentum representation
of the lattice theory. In order to make a Fourier transformation
to get the coordinate space representation,
it is useful to introduce the following bracket notation:
\begin{eqnarray}
&&~~~~~~\langle X |A \rangle =\frac{1}{N^{2}}\omega^{A\cdot X}
 \equiv \frac{1}{N^{2}}
 \omega^{a_0 x^0}\omega^{a_1 x^1}\omega^{a_2 x^2}\omega^{a_3 x^3}
\\
&&
\sum_{X\in \Z_N^4} \langle A | X \rangle \langle X | B \rangle
 =\delta_{AB},\quad 
\sum_{A\in \Z_N^4} \langle X | A \rangle \langle A | Y \rangle
=\delta_{XY},
\end{eqnarray}
where $X=(x^0,x^1,x^2,x^3)\in \Z_N^4$ is the coordinate space.
The matrix element $(U_{\mu})^{A}_{~B}$ is written
as $\langle A|U_{\mu}|B\rangle $ and its coordinate representation
is
\begin{eqnarray}
\langle X|U_{\mu}|Y\rangle
&=&\sum_{A,B} \langle X | A \rangle \sbra{A}U_{\mu}\sket{B}
  \langle B | Y \rangle \\
&=&\frac{1}{N^4}\sum_{A,B} \omega^{A\cdot X-B\cdot Y}
(U_{\mu})^{A}_{~B}.
\label{Umat}
\end{eqnarray}
Using the constraint (\ref{Uconst}),
we have $(U_\mu)^A_{~B}=\omega^{b_\mu} (U_\mu)^{A-B}_{~~~~~0}$
and (\ref{Umat}) can be calculated as
\begin{eqnarray}
\sbra{X}U_\mu\sket{Y}=
&=&\frac{1}{N^4}\sum_{A,B} \omega^{A\cdot X-B\cdot(Y-\hat\mu)}
(U_{\mu})^{A-B}_{~~~~~0}\\
&=&\frac{1}{N^4}\sum_{A'}(U_{\mu})^{A'}_{~~0}\,\,
\omega^{A'\cdot X} \sum_B \omega^{B\cdot(X+\hat\mu-Y)}\\
&=&U_\mu(X)\,\delta_{X+\hat\mu,Y},
\end{eqnarray}
where we have defined $U_\mu(X)\equiv\sum_A(U_\mu)^A_{~0}\,\omega^{A\cdot X}$
and used the relation
$\frac{1}{N^4}\sum_B\omega^{B\cdot(X+\hat\mu-Y)}=\delta_{X+\hat\mu,Y}$.
Similarly, we obtain
$\sbra{X}U_\mu^\dag\sket{Y}=U_\mu^\dag(Y)\delta_{X-\hat\mu,Y}$
for the hermitian conjugate of the matrix $U_\mu$.

Then the trace of $U_{\mu\nu}$ can be calculated as
\begin{eqnarray}
& &  \Tr (U_{\mu}U_{\nu}U_{\mu}^{\dagger}U_{\nu}^{\dagger})
\nonumber \\
& =&\sum_{X_{1},X_{2},X_{3},X_{4}}
\langle X_{1} | U_{\mu}| X_{2} \rangle \langle X_{2} | U_{\nu}| X_{3} \rangle
\langle X_{3} | U_{\mu}^{\dag}| X_{4} \rangle
\langle X_{4} | U_{\nu}^{\dag}| X_{1} \rangle \nonumber \\
&=&\sum_{X_{1},X_{2},X_{3},X_{4}}
U_{\mu}(X_{1})\delta_{X_{1}+\hat{\mu},X_{2}}
U_{\nu}(X_{2})\delta_{X_{2}+\hat{\nu},X_{3}}
 U_{\mu}^{\dag}(X_{4})\delta_{X_{3}-\hat{\mu},X_{4}}
U_{\nu}^{\dag}(X_{1})\delta_{X_{4}-\hat{\nu},X_{1}}
\nonumber \\
&=& 
\sum_{X}U_{\mu}(X)\,U_{\nu}(X+\wh{\mu})\,
U_{\mu}^{\dag}(X+\wh{\nu})\,
U_{\nu}^{\dag}(X),
\end{eqnarray}
which is the sum of the plaquettes over the space-time,
and hence the first trace in the (\ref{Uaction})
gives the usual action for the gauge field on the lattice.

The matter part is also performed similarly.
The matrix element of the fermion $\lambda^I$
in the coordinate space is given by
\begin{eqnarray}
\sbra{X}\lambda^I\sket{Y}=\lambda^I(X)\,\delta_{X,Y},
\end{eqnarray}
where we have used the constraint (\ref{con2}) and
defined $\lambda^I(X)=\sum_A(\lambda^I)^A_{~0}\,\omega^{A\cdot X}$.
The matrix $\Omega_\mu$ defined in (\ref{Ome}) acts as a shift matrix
\begin{eqnarray}
\sbra{X}\Omega_\mu\sket{Y}=\delta_{X+\hat\mu,Y},
\end{eqnarray}
in the coordinate space.
Then, the matter part of the action (\ref{Uaction}) becomes
\begin{eqnarray}
&&\Tr \left(\ol{\lambda}_I\gamma^{\mu}(U_{\mu}\lambda^I\Omega^{-1}
-U_{\mu}^{\dag}\lambda^I\Omega)\right)\nn\\
&=&\sum_X\ol\lambda_I(X)\gamma^\mu 
\left(U_\mu(X)\lambda^I(X+\wh\mu)-U_\mu^\dag(X)\lambda^I(X-\wh\mu)\right),
\end{eqnarray}
which is the standard (naive) fermion action for the lattice gauge theory.

A few comments are in order.
A similar approach to the matrix description of the lattice gauge
theory can be found in \cite{Kiku1,Neu,Kiku2}. In these papers,
the matrix variables are treated in the coordinate space
($\sbra{X}U_\mu\sket{Y}$ etc. in our notation).
The essential step in our discussion is the finite size regularization
of the infinite size matrices. We have replaced the $\Z$
valued matrix indices to $\Z_N$ valued ones. This is essentially
the same procedure in the dimensional deconstruction
\cite{Arkani1},
in which the regularization is obtained by
replacing a cylinder ($\R\times \R/\Z$) geometry 
with an orbifold ($\C/\Z_N$)\cite{Arkani2}.

\section{Discussion}
\label{discuss}

We have seen that two-dimensional and four-dimensional
QED are realized in string theory as the world-volume
theory on a probe $D1$-brane and $D3$-brane, respectively,
in the presence of $D9$-$\ol{D9}$ pairs in type IIB string theory.
In particular, the field theoretical methods developed in
QED${}_2$ have been efficiently used to study some
properties of the D-branes in this system.

Though we have concentrated on the $D9$-$\ol{D9}$ system
and its T-dual cousins, it should be possible to study
other unstable D-brane systems in a similar way.
It would be interesting to make more general investigation
on the unstable D-brane systems using the D-brane probes.

Our motivation to use a $D1$-brane as a probe
is not only that the world-sheet theory is a solvable field theory,
but also that it is mapped to a fundamental string under the S-duality
in type IIB string theory.
If we wish to study the $D9$-$\ol{D9}$ system in the
strong coupling region, it is natural to
take the $D1$-brane as a fundamental object.
This is quite analogous to the S-duality between type I string theory
and Heterotic $SO(32)$ string theory,
in which the field contents on the $D1$-brane
in type I string theory is the same as those
on the fundamental string in the Heterotic string theory.\cite{PoWi} \
The vertex operators for the ten dimensional
gauge fields which can be read
{}from (\ref{D9gauge}) are the analog of that for
the $SO(32)$ gauge field in the fermionic construction
of the Heterotic string.
However, unlike the type I $D1$-brane,
the field contents listed in table \ref{p=1} do not seem to lead to a
conformally invariant theory in a simple way even in the $n=1$ case.
It may be too naive to make such observation,
since we have taken the weak coupling
limit $g_s\ra 0$ to determine the action which may not give a good
description in the strong coupling region.
We hope our probe analysis will be helpful
to gain some insights to the strong coupling analysis
of the unstable D-brane system in the future study.

\section*{Acknowledgments}
\vskip2mm

We would like to thank our colleagues in Yukawa Institute for
Theoretical Physics for discussions and encouragement.
We are especially grateful to K. Fukaya, T. Onogi and T. Takimi
for useful discussions on the Schwinger model and the lattice
gauge theory.
S.S. also would like to thank Korea Institute
for Advanced Study, where a part of this work was done,
for hospitality and P. Yi and  Y. Kim for enjoyable discussion.
This work is supported by a Grant-in-Aid for the 21st Century COE
``Center for Diversity and Universality in Physics''.
The work of S.S. is partly supported by
Rikkyo University Special Fund for Research.

\end{document}